\documentclass[aps,twocolumn,showpacs]{revtex4}
\usepackage{graphicx}

\begin{document}
\title{Symbolic Stochastic Dynamical Systems Viewed as
Binary $N$-Step Markov Chains}
\author{O. V. Usatenko
\footnote[1]{usatenko@ire.kharkov.ua}, V. A. Yampol'skii}
\affiliation{A. Ya. Usikov Institute for Radiophysics and Electronics \\
Ukrainian Academy of Science, 12 Proskura Street, 61085 Kharkov,
Ukraine}
\author{K. E. Kechedzhy, S. S. Mel'nyk}
\affiliation{Department of Physics, Kharkov National University, 4
Svoboda Sq., Kharkov 61077, Ukraine}

\begin{abstract}
A theory of systems with long-range correlations based on the
consideration of \textit{binary N-step Markov chains} is
developed. In the model, the conditional probability that the
$i$-th symbol in the chain equals zero (or unity) is a linear
function of the number of unities among the preceding $N$ symbols.
The correlation and distribution functions as well as the variance
of number of symbols in the words of arbitrary length $L$ are
obtained analytically and numerically. A self-similarity of the
studied stochastic process is revealed and the similarity group
transformation of the chain parameters is presented. The diffusion
Fokker-Planck equation governing the distribution function of the
$L$-words is explored. If the persistent correlations are not
extremely strong, the distribution function is shown to be the
Gaussian with the variance being nonlinearly dependent on $L$. The
applicability of the developed theory to the coarse-grained
written and DNA texts is discussed.
\end{abstract}

\date{\today}


\maketitle

\section{Introduction}

The problem of systems with long-range spatial and/or
temporal correlations (LRCS) is one of the topics of
intensive research in modern physics, as well as in the
theory of dynamical systems and the theory of probability.
The LRC-systems are usually characterized by a complex
structure and contain a number of hierarchic objects as
their subsystems. The LRC-systems are the subject of study
in physics, biology, economics, linguistics, sociology,
geography, psychology, etc.~\cite{stan,prov,mant,kant}. At
the present time, there is no generally accepted theoretical
model that adequately describes the dynamical and
statistical properties of the LRC-systems. Attempts to
describe the behavior of the LRCS in the framework of the
Tsalis non-extensive thermodynamics~\cite{tsal,abe} were
undertaken in Ref.~\cite{den}. However, the non-extensive
thermodynamics is not well-grounded and requires the
construction of the additional models which could clarify
the properties of the LRC-systems.

One of the efficient methods to investigate the correlated
systems is based on a decomposition of the space of states
into a finite number of parts labelled by definite symbols.
This procedure referred to as coarse graining is accompanied
by the loss of short-range memory between states of system
but does not affect and does not damage its robust invariant
statistical properties on large scales. The most frequently
used method of the decomposition is based on the
introduction of two parts of the phase space. In other
words, it consists in mapping the two parts of states onto
two symbols, say 0 and 1. Thus, the problem is reduced to
investigating the statistical properties of the symbolic
binary sequences. This method is applicable for the
examination of both discrete and continuous systems.

One of the ways to get a correct insight into the nature of
correlations consists in an ability of constructing a
mathematical object (for example, a correlated sequence of
symbols) possessing the same statistical properties as the
initial system. There are many algorithms to generate
long-range correlated sequences: the inverse Fourier
transform~\cite{czir}, the expansion-modification Li
method~\cite{li}, the Voss procedure of consequent random
addition~\cite{voss}, the correlated Levy walks~\cite{shl},
etc.~\cite{czir}. We believe that, among the above-mentioned
methods, using the Markov chains is one of the most
important. We would like to demonstrate this statement in
the present paper.

In the following sections, the statistical properties of the
\textit{binary many-steps Markov chain} is examined. In
spite of the long-time history of studying the Markov
sequences (see, for example,~\cite{kant,nag,trib} and
references therein), the concrete expressions for the
variance of sums of random variables in such strings have
not yet been obtained. Our model operates with two
parameters governing the conditional probability of the
discrete Markov process, specifically with the memory length
$N$ and the correlation parameter $\mu$. The correlation and
distribution functions as well as the variance $D$ being
nonlinearly dependent on the length $L$ of a word are
derived analytically and calculated numerically. The
nonlinearity of the $D(L)$ function reflects the existence
of strong correlations in the system. The evolved theory is
applied to the coarse-grained written texts and
dictionaries, and to DNA strings as well.

Some preliminary results of this study were published in
Ref.~\cite{prl}.

\section{Formulation of the problem}

\subsection{Markov Processes}

Let us consider a homogeneous binary sequence of symbols,
$a_{i}=\{0,1\}$. To determine the $N$-\textit{step Markov
chain} we have to introduce the conditional probability
$P(a_{i}\mid a_{i-N},a_{i-N+1},\dots ,a_{i-1})$ of occurring
the definite symbol $a_i$ (for example, $a_i =0$) after
symbols $a_{i-N},a_{i-N+1},\dots ,a_{i-1}$. Thus, it is
necessary to define $2^{N}$ values of the $P$-function
corresponding to each possible configuration of the symbols
$a_{i-N},a_{i-N+1},\dots ,a_{i-1}$. We suppose that the
$P$-function has the form,
\[
P(a_{i}=0\mid a_{i-N},a_{i-N+1},\dots ,a_{i-1})
\]
\begin{equation}
=\frac{1}{N} \sum\limits_{k=1}^{N}f(a_{i-k},k).  \label{1}
\end{equation}
Such a relation corresponds to the additive influence of the
previous symbols on the generated one. The homogeneity of the
Markov chain is provided by the independence of the conditional
probability Eq.~(\ref{1}) of the index $i$.

It is reasonable to assume the function $f$ to be decreasing with
an increase of the distance $k$ between the symbols $a_{i-k}$ and
$a_{i}$ in the Markov chain. However, for the sake of simplicity
we consider here a step-like memory function $f(a_{i-k},k)$
independent of the second argument $k$. As a result, the model is
characterized by three parameters only, specifically by $f(0)$,
$f(1)$, and $N$:
\[
P(a_{i}=0\mid a_{i-N},a_{i-N+1},\dots ,a_{i-1})
\]
\begin{equation}
=\frac{1}{N} \sum\limits_{k=1}^{N}f(a_{i-k}).  \label{2}
\end{equation}
Note that the probability $P$ in Eq.~(\ref{2}) depends on the
numbers of symbols 0 and 1 in the $N$-word but is independent of
the arrangement of the elements $a_{i-k}$. We also suppose that
\begin{equation}
f(0)+f(1)=1.  \label{2a}
\end{equation}
This relation provides the statistical equality of the numbers of
symbols zero and unity in the Markov chain under consideration. In
other words, the chain is non-biased. Indeed, taking into account
Eqs.~(\ref{2}) and (\ref{2a}) and the sequence of equations,
\[
P(a_{i} = 1|a_{i-N},\dots ,a_{i-1})=1-P(a_{i}=0|a_{i-N},\dots
,a_{i-1})
\]
\begin{equation}
=\frac{1}{N}\sum\limits_{k=1}^{N}f(\tilde{a}_{i-N})= P(a_{i}=0\mid
\tilde{a} _{i-N},\dots ,\tilde{a}_{i-1}), \label{2b}
\end{equation}
one can see the symmetry with respect to interchange
$\tilde{a}_{i}\leftrightarrow a_{i}$ in the Markov chain.
Here $\tilde{a}_{i}$ is the symbol opposite to $a_{i}$,
$\tilde{a}_{i}=1-a_{i}$. Therefore, the probabilities of
occurring the words $(a_{1},\dots ,a_{L})$ and
$(\tilde{a}_{1},\dots ,\tilde{a}_{L})$ are equal to each
other for any word length $L$. At $L=1$ this yields equal
average probabilities that symbols $0$ and $1$ occur in the
chain.

Taking into account the symmetry of the conditional probability
$P$ with respect to a permutation of symbols $a_{i}$ (see
Eq.~(\ref{2})), we can simplify the notations and introduce the
conditional probability $p_{k}$ of occurring the symbol zero after
the $N$-word containing $k$ unities, e.g., after the word
$\underbrace{(11...1}_{k}\;\underbrace{00...0}_{N-k})$,
\[
p_{k}=P(a_{N+1}=0\mid \underbrace{11\dots
1}_{k}\;\underbrace{00\dots 0} _{N-k})
\]
\begin{equation}
=\frac{1}{2}+\mu (1-\frac{2k}{N}),  \label{14}
\end{equation}
with the correlation parameter $\mu $ being defined by the
relation
\begin{equation}
\mu =f(0)-\frac{1}{2}.  \label{3}
\end{equation}

We focus our attention on the region of $\mu $ determined by
the persistence inequality $0 < \mu <1/2$. In this case,
each of the symbols unity in the preceding N-word promotes
the birth of new symbol unity. Nevertheless, the major part
of our results is valid for the anti-persistent region
$-1/2<\mu <0$ as well.

A\hfill similar\hfill rule\hfill for\hfill the\hfill production\hfill
of\hfill an\hfill $N$-word \\
$(a_{1},\dots,a_{N})$ that follows after a word $(a_{0},a_1,\dots
,a_{N-1})$ was suggested in Ref.~\cite{kant}. However, the
conditional probability $p_k$ of occurring the symbols $a_N$ does
not depend on the previous ones in the model~\cite{kant}.

\subsection{Statistical characteristics of the chain}

In order to investigate the statistical properties of the Markov
chain, we consider the distribution $W_{L}(k)$ of the words of
definite length $L$ by the number $k$ of unities in them,
\begin{equation}
k_{i}(L)=\sum\limits_{l=1}^{L}a_{i+l},  \label{5}
\end{equation}
and the variance of $k$,
\begin{equation}
D(L)=\overline{k^{2}}-\overline{k}^{2},  \label{7}
\end{equation}
where
\begin{equation}
\overline{f(k)}=\sum\limits_{k=0}^{L}f(k)W_{L}(k).  \label{8}
\end{equation}
If $\mu =0,$ one arrives at the known result for the
non-correlated Brownian diffusion,
\begin{equation}
D(L)=L/4.  \label{6}
\end{equation}
We will show that the distribution function $W_{L}(k)$ for
the sequence determined by Eq.~(\ref{14}) (with nonzero but
not extremely close to 1/2 parameter $\mu $) is the Gaussian
with the variance $D(L)$ nonlinearly dependent on $L$.
However, at $\mu \rightarrow 1/2$ the distribution function
can differ from the Gaussian.

\subsection{Main equation}

For\hfill the\hfill stationary\hfill Markov\hfill
chain,\hfill the\hfill probability \\
$b(a_{1}a_{2}\dots a_{N})$ of occurring a certain word
$(a_{1},a_{2},\dots ,a_{N})$ satisfies the condition of
compatibility for the Chapmen-Kolmogorov equation (see, for
example, Ref.~\cite{gar}):
\[
b(a_{1}\dots a_{N})
\]
\begin{equation}
=\sum_{a=0,1}b(aa_{1}\dots a_{N-1})P(a_{N}\mid a,a_{1},\dots
,a_{N-1}).  \label{10}
\end{equation}
Thus, we have $2^{N}$ homogeneous algebraic equations for
the $2^{N}$ probabilities $b$ of occurring the $N$-words and
the normalization equation $\sum b=1$. In the case under
consideration, the set of equations can be substantially
simplified owing to the following statement.

\textbf{Proposition} $\spadesuit$: \textit{The probability
$b(a_{1}a_{2}\dots a_{N})$ depends on the number $k$ of unities in
the $N$-word only}, i.\ e., it is independent of the arrangement
of symbols in the word $(a_{1},a_{2},\dots ,a_{N})$.
\begin{figure}[h!]
{\includegraphics[width=0.45\textwidth,height=0.35\textwidth]{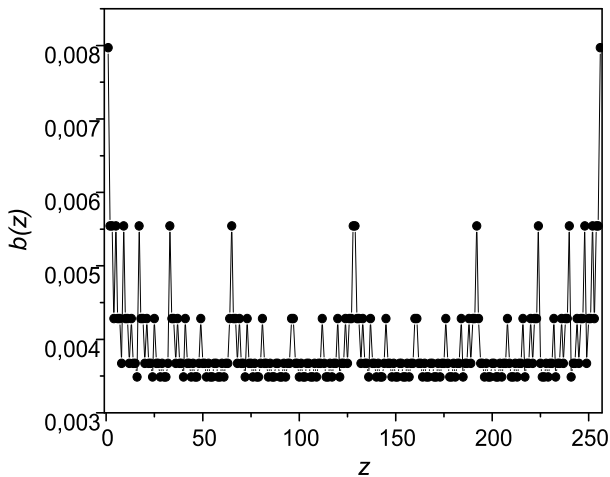}}
\caption{The probability $b$ of occurring a word $(a_1, a_2,
\dots , a_ N)$ vs its number $z$ expressed in the binary
code, $z=\sum_{i=1}^N a_i \cdot 2^{i-1}$, for $N=8$,
$\mu=0.4$.} \label{f1}
\end{figure}

This statement illustrated by Fig.~1 is valid owing to the
chosen simple model (\ref{2}), (\ref{14}) of the Markov
chain. It can be easily verified directly by substituting
the obtained below solution (\ref{b}) into the
set~(\ref{10}). Note that according to the Markov theorem,
Eqs.~(\ref{10}) do not have other solutions~\cite{kat}.

Proposition $\spadesuit$ leads to the very important
property of isotropy: any word $(a_{1},a_{2},\dots ,a_{L})$
appears with the same probability as the inverted one,
$(a_{L},a_{L-1},\dots ,a_{1})$.

Let\hfill us\hfill apply\hfill the\hfill set\hfill
of\hfill Eqs.~(\ref{10})\hfill to\hfill the\hfill word \\
$(\underbrace{11\dots 1}_{k}\;\underbrace{00\dots 0}_{N-k})$:
\[
b(\underbrace{11\dots 1}_{k}\;\underbrace{00\dots 0}_{N-k}) =b(0
\underbrace{11\dots 1}_{k}\;\underbrace{00\dots 0}_{N-k-1})p_{k}
\]
\begin{equation} \label{13}
+b(1\underbrace{11\dots 1}_{k}\;\underbrace{00\dots
0}_{N-k-1})p_{k+1}.
\end{equation}
This\hfill yields\hfill the\hfill recursion\hfill
relation\hfill for\hfill $b(k)=$ \\
$b(\underbrace{11...1}_{k}\; \underbrace{00...0}_{N-k})$,
\[
b(k)=\frac{1-p_{k-1}}{p_{k}}b(k-1)
\]
\begin{equation}
=\frac{N-2\mu (N-2k+2)}{N+2\mu (N-2k)} b(k-1).  \label{15}
\end{equation}
The probabilities $b(k)$ for $\mu>0$ satisfy the sequence of
inequalities,
\begin{equation}
b(0)=b(N)>b(1)=b(N-1)>...>b(N/2), \label{15b}
\end{equation}
which is the reflection of persistent properties for the
chain. At $\mu=0$ all probabilities are equal to each other.

The solution of Eq.~(\ref{10}) is
\begin{equation}\label{b}
b(k)=A\cdot \Gamma ( n+k) \Gamma ( n+N-k)
\end{equation}
with the parameter $n$ defined by
\begin{equation}\label{18a}
n= \frac{N(1-2\mu)}{4\mu}.
\end{equation}
The constant $A$ will be found below by normalizing the
distribution function. Its value is,
\begin{equation}
A=\frac{4^n }{2\sqrt{\pi}}
\frac{\Gamma(1/2+n)}{\Gamma(n)\Gamma(2n+N)}.\label{17a}
\end{equation}

\section{Distribution function of $L$-words}

In this section we investigate the statistical properties of
the Markov chain, specifically, the distribution of the
words of definite length $L$ by the number $k$ of unities.
The length $L$ can also be interpreted as the number of
jumps of some particle over an integer-valued 1-D lattice or
as the time of the diffusion imposed by the Markov chain
under consideration. The form of the distribution function
$W_{L}(k)$ depends, to a large extent, on the relation
between the word length $L$ and the memory length $N$.
Therefore, the first thing we will do is to examine the
simplest case $L = N$.

\subsection{Statistics of $N$-words}

The value $b(k)$ is the probability that an $N$-word contains $k$
unities with a \textit{definite} order of symbols $a_i$.
Therefore, the probability $W_{N}(k)$ that an $N$-word contains
$k$ unities with \textit{arbitrary} order of symbols $a_i$ is
$b(k)$ multiplied by the number $\mathrm{C}_{N}^{k}=N!/k!(N-k)!$
of different permutations of $k$ unities in the $N$-word,
\begin{equation}
W_{N}(k)=\text{C}_{N}^{k}b(k).  \label{19}
\end{equation}
Combining Eqs.~(\ref{b}) and (\ref{19}), we find the
distribution function,
\begin{equation}
W_{N}(k)= W_{N}(0)\text{C}_{N}^{k}\frac{\Gamma ( n+k) \Gamma (
n+N-k) }{\Gamma (n ) \Gamma (n+N)}.  \label{18}
\end{equation}
The normalization constant $W_{N}(0)$ can be obtained from the
equality $\sum\limits_{k=0}^{N}W_{N}(k)=1$,
\begin{equation}
W_N(0)=\frac{4^n }{2\sqrt{\pi}}
\frac{\Gamma(n+N)\Gamma(1/2+n)}{\Gamma(2n+N)}.\label{17}
\end{equation}
Comparing Eqs.~(\ref{b}), (\ref{19})-(\ref{17}), one can get
Eq.~(\ref{17a}) for the constant $A$ in Eq.~(\ref{b}).

Note that the distribution $W_{N}(k)$ is an even function of
the variable $\kappa =k-N/2$,
\begin{equation} W_{N}(N-k)=W_{N}(k).  \label{19b}
\end{equation}
This fact is a direct consequence of the above-mentioned
statistical equivalence of zeros and unities in the Markov
chain being considered. Let us analyze the distribution
function $W_{N}(k)$ for different relations between the
parameters $N$ and $\mu$.
\subsubsection{Limiting case of weak persistence, $n \gg 1$}

In the absence of correlations, $n \rightarrow \infty$,
Eq.~(\ref{18}) and the Stirling formula yield the Gaussian
distribution at $k,\, N,\, N-k \gg 1$. Given the persistence
is not too strong,
\begin{equation}\label{19c}
n \gg 1,
\end{equation}
one can also obtain the Gaussian form for the distribution
function,
\begin{equation}
W_{N}(k)=\frac{1}{\sqrt{2\pi D(N)}}\exp \left\{
-\frac{(k-N/2)^{2}}{2D(N)} \right\} ,  \label{27}
\end{equation}
with the $\mu$-dependent variance,
\begin{equation}
D(N)=\frac{N(N+2n)}{8n}=\frac{N}{4(1-2\mu )}.  \label{28}
\end{equation}
Equation (\ref{27}) says that the $N$-words with equal
numbers of zeros and unities, $k=N/2$, are most probable.
Note that the persistence results in an increase of the
variance $D(N)$ with respect to its value $N/4$ at $\mu =0$.
In other words, the persistence is conductive to the
intensification of the diffusion. Inequality $n \gg 1$ gives
$D(N) \ll N^{2}$. Therefore, despite the increase of $D(N)$,
the fluctuations of $(k-N/2)$ of the order of $N$ are
exponentially small.
\subsubsection{Intermediate case, $n \gtrsim 1$}

If the parameter $n$ is an integer of the order of unity, the
distribution function $W_{N}(k)$ is a polynomial of degree
$2(n-1)$. In particular, at $n=1$, the function $W_{N}(k)$ is
constant,
\begin{equation}
W_{N}(k)=\frac{1}{N+1}.  \label{24}
\end{equation}
At $n\neq 1,$ $W_{N}(k)$ has a maximum in the middle of the
interval $[0,N]$.
\subsubsection{Limiting case of strong persistence}
If the parameter $n$ satisfies the inequality,
\begin{equation}\label{24a}
n \ll \ln^{-1}N,
\end{equation}
one can neglect the parameter $n$ in the arguments of the
functions $\Gamma (n+k)$, $\Gamma (n+N)$, and $\Gamma
(n+N-k)$ in Eq.~(\ref{18}). In this case, the distribution
function $W_{N}(k)$ assumes its maximal values at $k=0$ and
$k=N$,
\begin{equation}
W_{N}(1)=W_{N}(0)\frac{nN}{N-1} \ll W_{N}(0). \label{20}
\end{equation}
Formula (\ref{20}) describes the sharply decreasing
$W_{N}(k)$ as $k$ varies from $0$ to $1$ (and from $N$ to
$N-1$). Then, at $1<k<N/2$, the function $W_{N}(k)$
decreases more slowly with an increase in $k$,
\begin{equation}
W_{N}(k)=W_{N}(0)\frac{nN}{k(N-k)}.  \label{21}
\end{equation}
At $k=N/2,$ the probability $W_{N}(k)$ achieves its minimal value,
\begin{equation}
W_{N}\left(\frac{N}{2}\right)= W_{N}(0)\frac{4n}{N}. \label{22}
\end{equation}

It follows from normalization (\ref{17}) that the values
$W_{N}(0)=W_N (N)$ are approximatively equal to $1/2$.
Neglecting the terms of the order of $n^2$, one gets
\begin{equation}
W_{N}(0)=\frac{1}{2} ( 1 - n \ln N ). \label{22a}
\end{equation}
In the straightforward calculation using Eqs. (\ref{7}) and
(\ref{21}) the variance $D$ is
\begin{equation}
D(N)=\frac{N^2}{4} -\frac{nN(N-1)}{2}. \label{22b}
\end{equation}

Thus, the variance $D(N)$ is equal to $N^2 /2$ in the leading
approximation in the parameter $n$. This fact has a simple
explanation. The probability of occurrence the $N$-word containing
$N$ unities is approximatively equal to $1/2$. So, the relations
$\overline{k^{2}} \approx N^2/2 $ and $\overline{k}^{2}=N^2/4$
give (\ref{22b}). The case of strong persistence corresponds to
the so-called ballistic regime of diffusion: if we chose randomly
some symbol $a_i$ in the sequence, it will be surrounded by the
same symbols with the probability close to unity.

The evolution of the distribution function $W_N(k)$ from the
Gaussian form to the inverse one with a decrease of the
parameter $n$ is shown in Fig.~2. In the interval $\ln^{-1}N
< n < 1 $ the curve $W_{N}(k)$ is concave and the maximum of
function $W_{N}(k)$ inverts into minimum. At $N \gg 1 $ and
$\ln^{-1}N < n < 1 $, the curve remains a smooth function of
its argument $k$ as shown by curve with $n=0.5$ in Fig.~2.
Below, we will not consider this relatively narrow region of
the change in the parameter $n$.

Formulas (\ref{27}), (\ref{28}), (\ref{21}), (\ref{22a}) and
(\ref{22b}) describe the statistical properties of $L$-words for
the fixed ''diffusion time'' $L=N$. It is necessary to examine the
distribution function $W_{L}(k)$ for the general situation, $L\neq
N$. We start the analysis with $L<N$.
\begin{figure}[h!]
{\includegraphics[width=0.45\textwidth,height=0.35\textwidth]{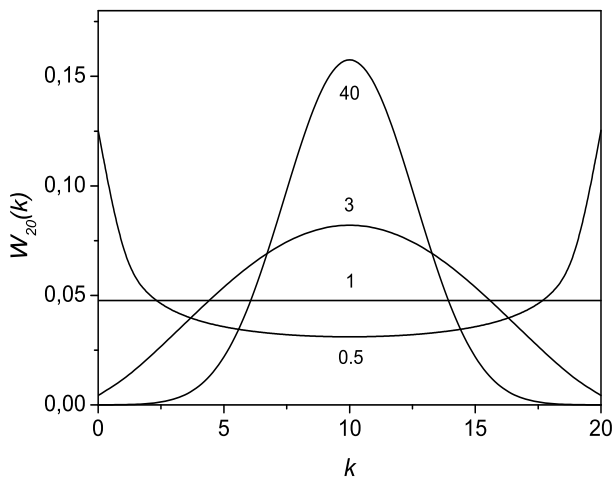}}
\caption{The distribution function $W_N(k)$ for $N$=20 and
different values of the parameter $n$ shown near the curves.}
\label{f2}
\end{figure}
\subsection{Statistics of $L$-words with $L<N$}

\subsubsection{Distribution function $W_{L}(k)$}

The distribution function $W_{L}(k)$ at $L<N$ can be given as
\begin{equation}
W_{L}(k)=\sum\limits_{i=k}^{k+N-L}b(i)\text{C}_{L}^{k}\text{C}_{N-L}^{i-k}.
\label{29}
\end{equation}
This equation follows from the consideration of $N$-words
consisting of two parts,
\begin{equation}
(\underbrace{a_{1},\dots ,a_{N-L},}_{i-k\text{
unities}}\;\underbrace{a_{N-L+1},\dots ,a_{N}}_{k \text{
unities}}). \label{29b}
\end{equation}
The total number of unities in this word is $i$. The right-hand
part of the word ($L$-sub-word) contains $k$ unities. The
remaining ($i-k$) unities are situated within the left-hand part
of the word (within $(N-L)$-sub-word). The multiplier
$\mathrm{C}_{L}^{k}\mathrm{C}_{N-L}^{i-k}$ in Eq.~(\ref{29}) takes
into account all possible permutations of the symbols ''1'' within
the $N$-word on condition that the $L$-sub-word always contains
$k$ unities. Then we perform the summation over all possible
values of the number $i$. Note that Eq.~(\ref{29}) is a direct
consequence of the proposition $\spadesuit$ formulated in
Subsec.~C of the previous section.

The straightforward summation in Eq.~(\ref{29}) yields the
following formula that is valid at any value of the parameter $n$:
\begin{equation}\label{W(L)}
W_L(k)=W_L(0)\text{C}_{L}^{k}\frac{\Gamma(n+k) \Gamma
(n+L-k)}{\Gamma (n) \Gamma (n+L)}
\end{equation}
where
\begin{equation}\label{W(0)}
W_L(0)=\frac{4^n }{2\sqrt{\pi}}\frac{\Gamma(1/2+n)\Gamma
(n+L)}{\Gamma (2n+L)}.
\end{equation}

It is of interest to note that the parameter of persistence $\mu$
and the memory length $N$ are presented in Eqs.~(\ref{W(L)}),
(\ref{W(0)}) via the parameter $n$ only. This means that the
statistical properties of the $L$-words with $L<N$ are defined by
this single "combined" parameter.

In the limiting case of weak persistence, $n\gg 1$, at $k,\;L-k
\gg 1$, Eq.~(\ref{W(L)}) along with the Stirling formula give the
Gaussian distribution function,
\begin{equation}
W_{L}(k)=\frac{1}{\sqrt{2\pi D(L)}}\exp \left\{
-\frac{(k-L/2)^{2}}{2D(L)} \right\}  \label{31}
\end{equation}
with the variance $D(L)$,
\begin{equation}
D(L)=\frac{L}{4}\left(1+\frac{L}{2n}\right)=
\frac{L}{4}\left[1+\frac{2\mu L}{N(1-2\mu )}\right]. \label{32}
\end{equation}

In the case of strong persistence (\ref{24a}), the
asymptotic expression for the distribution function
Eq.~(\ref{W(L)}) can be written as
\begin{equation} \label{45f}
W_{L}(k)=W_{L}(0)\frac{nL}{k(L-k)}, \,\,\, k\neq 0,\,\, k\neq L,
\end{equation}
\begin{equation}
W_{L}(0)=W_{L}(L)=\frac{1}{2} ( 1 - n \ln L ). \label{45b}
\end{equation}
Both the distribution $W_{L}(k)$ (\ref{45f}) and the
function $W_{N}(k)$ (\ref{21}) has a concave form. The
former assumes the maximal value (\ref{45b}) at the edges of
the interval $[0, L]$ and has a minimum at $k=L/2$.

\subsubsection{Variance $D(L)$}

Using the definition Eq.~(\ref{7}) and the distribution function
Eq.~(\ref{W(L)}) one can obtain a very simple formula for the
variance $D(L)$,
\begin{equation}\label{D(L)}
D(L)=\frac{L}{4}[1+m(L-1)],
\end{equation}
with
\begin{equation}\label{m}
m=\frac{1}{1+2n}= \frac{2\mu}{N-2\mu(N-1)}.
\end{equation}
Eq.~(\ref{D(L)}) shows that the vari\-ance $D(L)$ obeys the
pa\-ra\-bo\-lic law independently of the correlation strength in
the Markov chain.

In the case of weak persistence, at $n\gg 1$, we obtain the
asymptotics Eq.~(\ref{32}). It allows one to analyze the
behavior of the variance $D(L)$ with an increase in the
``diffusion time'' $L$. At small $mL \ll 1$, the dependence
$D(L)$ follows the classical law of the Brownian diffusion,
$D(L)\approx L/4$. Then, at $mL\sim 1$, the function $D(L)$
becomes super-linear.

For the case of strong persistence, $n \ll 1$,
Eq.~(\ref{D(L)}) gives the asymptotics,
\begin{equation}
D(L)=\frac{L^2}{4} - \frac{nL(L-1)}{2}. \label{45c}
\end{equation}
The ballistic regime of diffusion leads to the quadratic law
of the $D(L)$ dependence in the zero approximation in the
parameter $n \ll 1$.

The unusual behavior of the variance $D(L)$ raises an issue
as to what particular type of the diffusion equation
corresponds to the nonlinear dependence $D(L)$ in
Eq.~(\ref{32}). In the following subsection, when solving
this problem, we will obtain the conditional probability
$p^{(0)}$ of occurring the symbol zero after a given
$L$-word with $L<N$. The ability to find $p^{(0)}$, with
some reduced information about the preceding symbols being
available, is very important for the study of the
self-similarity of the Markov chain (see Subsubsec.~4 of
this Subsection).

\subsubsection{Generalized diffusion equation at $L<N$, $n \gg 1$}

It is quite obvious that the distribution $W_{L}(k)$ satisfies the
equation
\begin{equation}
W_{L+1}(k)=W_{L}(k)p^{(0)}(k)+W_{L}(k-1)p^{(1)}(k-1).  \label{33}
\end{equation}
Here $p^{(0)}(k)$ is the probability of occurring ''0'' after an
average-statistical $L$-word containing $k$ unities and
$p^{(1)}(k-1)$ is the probability of occurring ''1'' after an
$L$-word containing $(k-1)$ unities. At $L<N$, the probability
$p^{(0)}(k)$ can be written as
\begin{equation}
p^{(0)}(k)=\frac{1}{W_L(k)}
\sum\limits_{i=k}^{k+N-L}p_{i}b(i)\mathrm{C}_{L}^{k}\mathrm{C}_{N-L}^{i-k}.
\label{34}
\end{equation}
The product $b(i)\mathrm{C}_{L}^{k}\mathrm{C}_{N-L}^{i-k}$ in this
formula represents the conditional probability of occurring the
$N$-word containing $i$ unities, the right-hand part of which, the
$L$-sub-word, contains $k$ unities (compare with Eqs.~(\ref{29}),
(\ref{29b})).

The product $b(i)\mathrm{C}_{N-L}^{i-k}$ in Eq.~(\ref{34})
is a sharp function of $i$ with a maximum at some point
$i=i_0$ whereas $p_{i}$ obeys the linear law (\ref{14}).
This implies that $p_{i}$ can be factored out of the
summation sign being taken at point $i=i_0$. The
asymptotical calculation shows that point $i_0$ is described
by the equation,
\begin{equation}
i_{0}=\frac{N}{2}-\frac{L/2}{1-2\mu (1-L/N)}\left(
1-\frac{2k}{L}\right). \label{35}
\end{equation}
Expression (\ref{14}) taken at point $i_0$ gives the desired
formula for $p^{(0)}$ because
\begin{equation}
\sum\limits_{i=k}^{k+N-L}b(i)\mathrm{C}_{L}^{k}\mathrm{C}_{N-L}^{i-k}
\end{equation}
is obviously equal to $W_L(k)$. Thus, we have
\begin{equation}
p^{(0)}(k)=\frac{1}{2}+\frac{\mu L}{N-2\mu (N-L)}\left(
1-\frac{2k}{L}\right). \label{36}
\end{equation}

Let us consider a very important point relating to
Eq.~(\ref{35}). If the concentration of unities in the
right-hand part of the word (\ref{29b}) is higher than
$1/2$, $k/L >1/2$, then the most probable concentration
$(i_0-k)/(N-L)$ of unities in the left-hand part of this
word is likewise increased, $(i_0-k)/(N-L)>1/2$. At the same
time, the concentration $(i_0-k)/(N-L)$ is less than $k/L$,
\begin{equation}\label{36b}
\frac{1}{2} <\frac{i_0-k}{N-L}<\frac{k}{L}.
\end{equation}
This implies that the increased concentration of unities in
the $L$-words is necessarily accompanied by the existence of
a certain tail with an increased concentration of unities as
well. Such a phenomenon is referred by us as the
\textit{macro-persistence}. An analysis performed in the
following section will indicate that the correlation length
$l_c$ of this tail is $\gamma N $ with $\gamma \geq 1$
dependent on the parameter $\mu$ only. It is evident from
the above-mentioned property of the isotropy of the Markov
chain that there are two correlation tails from both sides
of the $L$-word.

Note that the distribution $W_L(k)$ is a smooth function of
arguments $k$ and $L$ near its maximum in the case of weak
persistence and $k, L-k\gg 1$. By going over to the continuous
limit in Eq.~(\ref{33}) and using Eq.~(\ref{36}) with the relation
$p^{(1)}(k-1)=1-p^{(0)}(k-1)$, we obtain the diffusion
Fokker-Planck equation for the correlated Markov process,
\begin{equation}
\frac{\partial W}{\partial L}=\frac{1}{8}\frac{\partial
^{2}W}{\partial \kappa ^{2}}-\eta(L)\frac{\partial }{\partial
\kappa }( \kappa W), \label{39}
\end{equation}
where $\kappa =k-L/2$ and
\begin{equation}\label{39b}
\eta(L)=\frac{2\mu}{(1-2\mu )N+2\mu L}.
\end{equation}
Equation (\ref{39}) has a solution of the Gaussian form
Eq.~(\ref{31}) with the variance $D(L)$ satisfying the ordinary
differential equation,
\begin{equation}
\frac{\mathrm{d}D}{\mathrm{d}L}=\frac{1}{4}+2\eta(L)D. \label{40}
\end{equation}
Its solution, given the boundary condition $D(0)=0$,
coincides with (\ref {32}).

\subsubsection{Self-similarity of the persistent Brownian
diffusion}

In this subsection, we point to one of the most interesting
properties of the Markov chain being considered, namely, its
self-similarity. Let us reduce the $N$-step Markov sequence
by regularly (or randomly) removing some symbols and
introduce the decimation parameter $\lambda$,
\begin{equation}
\lambda =N^{\ast }/N \leq 1.  \label{41}
\end{equation}
Here $N^{\ast }$ is a renormalized memory length for the reduced
$N^{\ast }$-step Markov chain. According to Eq.~(\ref{36}), the
conditional probability $p_{k}^{\ast }$ of occurring the symbol
zero after $k$ unities among the preceding $N^{\ast }$ symbols is
described by the formula,
\begin{equation}
p_{k}^{\ast }=\frac{1}{2}+\mu ^{\ast }\left( 1-\frac{2k}{N^{\ast
}}\right), \label{42}
\end{equation}
with
\begin{equation}
\mu ^{\ast }=\mu \frac{\lambda }{1-2\mu (1-\lambda )}.  \label{43}
\end{equation}
The comparison between Eqs.~(\ref{14}) and (\ref{42}) shows
that the reduced chain possesses the same statistical
properties as the initial one but it is characterized by the
renormalized parameters ($N^{\ast }$, $\mu ^{\ast }$)
instead of ($N$, $\mu $). Thus, Eqs.~(\ref{41}) and
(\ref{43}) determine the one-parametrical renormalization of
the parameters of the stochastic process defined by
Eq.~(\ref{14}).

The astonishing property of the reduced sequence consists in that
\textit{the variance $D^{\ast }(L)$ is invariant with respect to
the one-parametric decimation transformation} (\ref{41}),
(\ref{43}). In other words, it coincides with the function $D(L)$
for the initial Markov chain:
\begin{equation} \label{44}
D^{\ast }(L) = \frac{L}{4}[1+m ^{\ast } (L-1)] = D(L), \qquad
L<N^{\ast }.
\end{equation}
Indeed, according to Eqs.~(\ref{41}), (\ref{43}), the renormalized
parameter $m ^{\ast }=2\mu ^{\ast}/[N^{\ast} - 2\mu ^{\ast}
(N^{\ast} -1)]$ of the reduced sequence coincides exactly with the
parameter $m =2\mu/[N - 2\mu (N-1)]$ of the initial Markov chain.
Since the shape of the function $W_L(k)$ Eq.~(\ref{W(L)}) is
defined by the invariant parameter $n=n^{\ast}$, the distribution
$W_L(k)$ is also invariant with respect to the decimation
transformation.

The transformation ($N$, $\mu $) $\rightarrow$ ($N^{\ast }$,
$\mu ^{\ast }$) (\ref{41}), (\ref{43}) possesses the
properties of semi-group, i.\ e., the composition of
transformations ($N$, $\mu $) $\rightarrow$ ($N^{\ast }$,
$\mu ^{\ast }$)  and ($N^{\ast }$, $\mu ^{\ast }$)
$\rightarrow$ ($N^{\ast \ast }$, $\mu ^{\ast \ast}$) with
transformation parameters $\lambda_1$ and $\lambda_2$ is
likewise the transformation from the same semi-group, ($N$,
$\mu$) $\rightarrow$ ($N^{\ast \ast }$, $\mu ^{\ast \ast}$),
with parameter $\lambda = \lambda_1 \lambda_2$.

The invariance of the function $D(L)$ at $L<N$ was referred
to by us as the phenomenon of \textit{self-similarity}. It
is demonstrated in Fig.~3 and is accordingly discussed
below, in Sec.~IV A.

It is interesting to note that the property of
self-similarity is valid for any strength of the
persistency. Indeed, the result Eq.~(\ref{36}) can be
obtained directly from Eqs.~(\ref{b})-(\ref{17a}), and
(\ref{34}) not only for $n\gg 1$ but also for the arbitrary
value of $n$.

 \protect\begin{figure}[h!]
{\includegraphics[width=0.45\textwidth,height=0.35\textwidth]{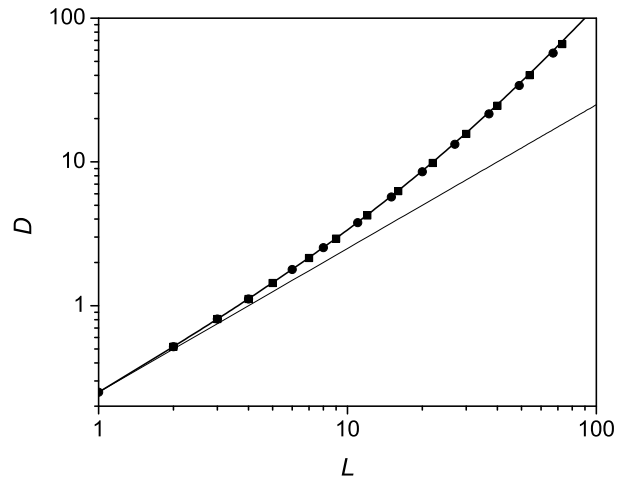}}
\caption{The dependence of the variance $D$ on the tuple length
$L$ for the generated sequence with $N=100$ and $\mu=0.4$ (solid
line) and for the decimated sequences (the parameter of decimation
$\lambda =0.5$). Squares and circles correspond to the stochastic
and deterministic reduction, respectively. The thin solid line
describes the non-correlated Brownian diffusion, $D(L)=L/4$.}
\label{f3}
\end{figure}

\subsection{Long-range diffusion, $L>N$}

Unfortunately, the very useful proposition $\spadesuit$ is
valid for the words of the length $L\leq N$ only and is not
applicable to the analysis of the long words with $L>N$.
Therefore, investigating the statistical properties of the
long words represents a rather challenging combinatorial
problem and requires new physical approaches for its
simplification. Thus, we start this subsection by analyzing
the correlation properties of the long words ($L>N$) in the
Markov chains with $N\gg 1$. The two first subsubsections of
this subsection mainly deal with the case of relatively weak
correlations, $n \gg 1$.

\subsubsection{Correlation length at weak persistence}

Let us rewrite Eq.~(\ref{14}) in the form,
\begin{equation}
<a_{i+1}>=\frac{1}{2}+\mu \left(
\frac{2}{N}\sum_{k=i-N+1}^{i}<a_{k}>-1 \right).  \label{46}
\end{equation}
The angle brackets denote the averaging of the density of
unities in some region of the Markov chain for its definite
realization. The averaging is performed over distances much
greater than unity but far less than the memory length $N$
and correlation length $l_c$ (see Eq.~(\ref{50b}) below).
Note that this averaging differs from the statistical
averaging over the ensemble of realizations of the Markov
chain denoted by the bar in Eqs.~(\ref{7}) and (\ref{8}).
Equation (\ref{46}) is a relationship between the average
densities of unities in two different macroscopic regions of
the Markov chain, namely, in the vicinity of $(i+1)$-th
element and in the region $(i-N,\,\,i)$. Such an approach is
similar to the mean field approximation in the theory of the
phase transitions and is asymptotically exact at
$N\rightarrow \infty$. In the continuous limit,
Eq.~(\ref{46}) can be rewritten in the integral form,
\begin{equation}
<a(i)>=\frac{1}{2}+\mu \left(
\frac{2}{N}\int_{i-N}^{i}<a(k)>\textrm{d}k-1\right). \label{47}
\end{equation}
It has the obvious solution,
\begin{equation}
<a(i)-\frac{1}{2}>=<a(0)-\frac{1}{2}>\exp \left(-i/\gamma
N\right),  \label{49}
\end{equation}
where the parameter $\gamma $ is determined by the relation,
\begin{equation}
\gamma \left( \exp \left( \frac{1}{\gamma }\right) -1\right)
=\frac{1}{2\mu}.  \label{50}
\end{equation}
A unique solution $\gamma $ of the last equation is an increasing
function of $\mu \in(0, 1/2)$.

Formula (\ref{49}) shows that any fluctuation (the difference
between $<a(i)>$ and the equilibrium value of
$\overline{a_i}=1/2$) is exponentially damped at distances of the
order of the \textit{correlation length} $l_c$,
\begin{equation}\label{50b}
l_{c}=\gamma N.
\end{equation}
Law (\ref{49}) describes the phenomenon of the
\textit{persistent macroscopic correlations} discussed in
the previous subsection. This phenomenon is governed by both
parameters, $N$ and $\mu$. According to Eqs.~(\ref{50}),
(\ref{50b}), the correlation length $l_c$ grows as $\gamma=
1/4 \delta$ with an increase in $\mu$ (at $\mu \rightarrow
1/2$) until the inequality $\delta \gg 1/N$ is satisfied.
Here
\begin{equation}\label{delta}
\delta = 1/2-\mu.
\end{equation}
Let us note that the inequality $\delta \gg 1/N$ defining the
regime of weak persistence can be rewritten in terms of $\gamma$,
$\gamma \ll N/4$. At $\delta \approx 1/N$, the correlation length
$l_c$ achieves its maximum value $N^2/4$. With the following
increase of $\mu$, the diffusion goes to the regime of strongly
correlated diffusion that will be discussed in Subsubsec 3 of this
Subsection.

At $\mu \rightarrow 0$, the macro-persistence is broken and the
correlation length tends to zero.

\subsubsection{Correlation function at weak persistence}

Using the studied correlation properties of the Markov
sequence and some heuristic reasons, one can obtain the
correlation function ${\cal K}(r)$ being defined as,
\begin{equation}
{\cal K}(r)=\overline{a_{i}a_{i+r}}-\overline{a_{i}}^2, \label{51}
\end{equation}
and then the variance $D(L)$. Comparing Eq.~(\ref{51}) with
Eqs.~(\ref{5}), (\ref{7}) and taking into account the
property of sequence, $\overline{a_{i}}=1/2$, it is easy to
derive the general relationship between functions ${\cal
K}(r)$ and $D(L)$,
\begin{equation} \label{51c}
D(L)=\frac{L^{2}}{4}+4\sum_{i=1}^{L-1}\sum_{r=1}^{L-i}{\cal K}(r).
\end{equation}
Considering (\ref{51c}) as an equation with respect to ${\cal
K}(r)$, one can find its solution,
\[
{\cal K}(1) = \frac{1}{2}D(2)-\frac{1}{4}, \quad {\cal K}(2) =
\frac{1}{2}D(3)-D(2)+\frac{1}{8},
\]
\begin{equation}\label{51e}
{\cal K}(r)=\frac{1}{2}\left[D(r+1) -2D(r) +D(r-1)\right], \quad
r\geq 3.
\end{equation}
This solution has a very simple form in the continuous limit,
\begin{equation}\label{51f}
{\cal K}(r) = \frac{1}{2}\frac{{\textrm d}^2 D(r)}{{\textrm
d}r^2}.
\end{equation}

\begin{figure}[h!]
{\includegraphics[width=0.45\textwidth,height=0.35\textwidth]{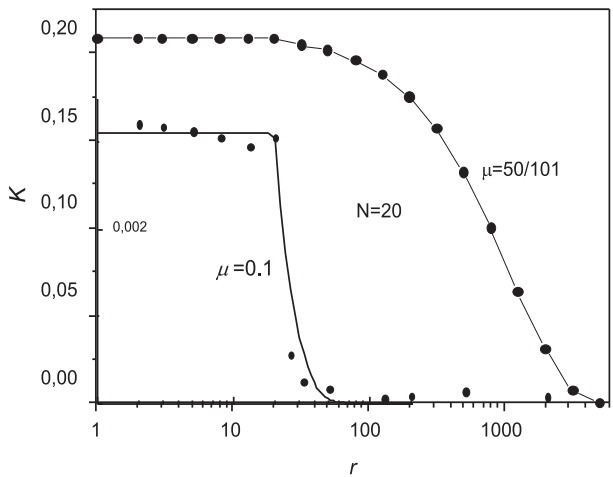}}
\caption{The dependence of the correlation function $K$ on
the distance $r$ between the symbols for the sequence with
$N=20$. The dots correspond to the generated sequence with
$\mu=0.1$  and $\mu=50/101$. The lower line is analytical
result (\ref{55}) with $l_c=\gamma N$ and $\gamma=0.38$.}
\label{f4}
\end{figure}

Equations~(\ref{51e}) and (\ref{D(L)}) give the correlation
function at $r<N$, $n\gg 1$,
\[
{\cal K}(r)=C_{r}m,
\]
with
\[
C_{1}=1/2, \qquad C_{2}=1/8, \qquad C_{3\leq r\leq N}=1/4,
\]
and $m$ determined by Eq.~(\ref{m}). In the continuous
approximation, the correlation function is described by the
formula,
\begin{equation}
{\cal K}(r)=\frac{m}{4 }, \qquad r \leq N. \label{54b}
\end{equation}
The independence of the correlation function of $r$ at $r<N$
results from our choice of the conditional probability in the
simplest form (\ref{14}). At $r>N$, the function ${\cal K}(r)$
should decrease because of the loss of memory. Therefore, using
Eqs.~(\ref{49}) and (\ref{50b}), let us prolongate the correlator
${\cal K}(r)$ as the exponentially decreasing function at $r>N$,
\begin{equation}
{\cal K}(r)=\frac{m}{4}\cases {1,\;\qquad \;\;\;\;\;\qquad r\leq
N, \cr \exp \left(-\frac{r-N}{l_{c}}\right ), \;\;r>N.} \label{55}
\end{equation}
The lower curve in Fig.~\ref{f4} presents the plot of the
correlation function at $\mu =0.1$.

According to Eqs.~(\ref{51f}), (\ref{55}), the variance
$D(L)$ can be written as
\begin{equation} \label{56}
D(L)=\frac{L}{4}\left(1+m F(L)\right)
\end{equation}
with
\begin{equation} \label{57}
F(L)= \cases {L,   \qquad \qquad \qquad \qquad \qquad \qquad L<N,
\cr2(1+ \gamma)N - (1+2\gamma ) \frac{N^2}{L} \cr -
2\gamma^{2}\frac{N^2}{L} \left[1-\exp \left(
-\frac{L-N}{l_c}\right) \right], \, \, L>N.}
\end{equation}
\begin{figure}[t!]
{\includegraphics[width=0.45\textwidth,height=0.35\textwidth]{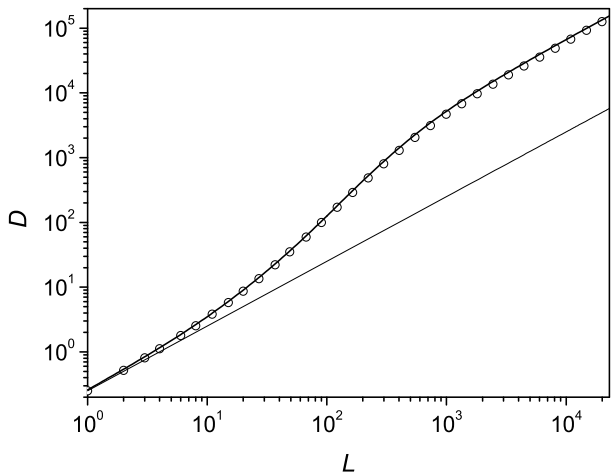}}
\caption{The numerical simulation of the dependence $D(L)$ for the
generated sequence with $N=100$ and $\mu=0.4$ (circles). The solid
line is the plot of function Eq.~(\ref{56}) with the same values
of $N$ and $\mu$.} \label{f5}
\end{figure}

As an illustration of the result Eq.~(\ref{56}), we present the
plot of $D(L)$ for $N=100$ and $\mu=0.4$ by the solid line in
Fig.~\ref{f5}. The straight line in the figure corresponds to the
dependence $D(L)=L/4$ for the usual Brownian diffusion without
correlations (for $\mu=0$). It is clearly seen that the plot of
variance (\ref{56}) contains two qualitatively different portions.
One of them, at $L\lesssim N$, is the super-linear curve that
moves away from the line $D=L/4$ with an increase of $L$ as a
result of the persistence. For $L\gg N$, the curve $D(L)$ achieves
the linear asymptotics,
\begin{equation}\label{58}
D(L)\cong \frac{L}{4}\left ( 1+ \frac{4\mu
(1+\gamma)}{1-2\mu}\right).
\end{equation}
This phenomenon can be interpreted as a result of the
diffusion in which every \textit{independent} step $\sim
\sqrt{D(L)}$ of wandering represents a path traversed by a
particle during the characteristic ``fluctuating time'' $L
\sim (N+l_c)$. Since these steps of wandering are
quasi-independent, the distribution function $W_L(k)$ is the
Gaussian. Thus, in the case of relatively weak persistence,
$n \gg 1$, $W_L(k)$ is the Gaussian not only at $L<N$ (see
Eq.~(\ref{31})) but also for $L>N, \, l_c$.

Note that the above-mentioned property of the self-similarity is
valid only at the portion $L<N$ of the curve $D(L)$. Since the
decimation procedure leads to the decrease of the parameter $\mu$
(see Eq.~(\ref{43})), the plot of asymptotics (\ref{58}) for the
reduced sequence at $L\gg N^{\ast}$ goes below the $D(L)$ plot for
the initial chain.

\subsubsection{Statistics of the $L$-words for the case of strong
persistence, $n \ll \ln^{-1}N $}

In this subsection, we study the statistical properties of
long words ($L>N$) in the sequences of symbols with strong
correlations. It is convenient to rewrite formula (\ref{14})
for the conditional probability of occurring the symbol zero
after the $N$-word containing $k$ unities in the form,
\begin{equation}
p_{\nu}=\delta+2\mu \frac{\nu}{N},  \label{58b}
\end{equation}
where $\nu$ is the number of zeros in the precedent
$N$-word, $\nu=N-k$.

In the case of strong persistence, $n \ll \ln^{-1}N $, the
parameter $\delta =1/2-\mu$ is much smaller than $1/N$.
Therefore, the probability $p_{\nu}$ can be written as
\begin{equation}
p_{\nu} \approx \cases {\delta,\;\qquad \,\,\,\,\,\,\,\,\,\,
\nu=0,  \cr \nu /N, \;\qquad \, \,\nu \neq0,\, \,\, \nu \neq
N, \cr 1-\delta, \;\qquad \, \nu=N.} \label{58d}
\end{equation}
It is seen that the probability of occurring the symbol zero
after the $N$-word which contains only unities ($\nu =0$)
represents very small value $\delta$ and it increases
significantly if $\nu \neq 0$.  This situation differs
drastically from the case of weak persistency. At $n \gg 1$,
the parameter $\delta$ exceeds noticeably the value $1/N$,
and the probability $p_\nu$ does not actually change with an
increase in the number of zeros in the preceding $N$-word.

The analysis of the symbol generation process in the Markov
chain in the case of strong persistence gives the following
picture of the fluctuations. There exist the entire portions
of the chain consisting of the same symbols, say unities.
The characteristic length of such portions is $1/\delta \gg
N$. These portions are separated by one or more symbols
zero. The number of such packets of the same symbols in one
fluctuation zone is about $N$. Thus, the characteristic
correlation distance at which the $N$-word containing the
same symbols converts into the $N$-word with $\nu =N/2$ is
about $N/\delta$,
\begin{equation}
l_c\approx \frac{N}{\delta}. \label{58e}
\end{equation}

\begin{figure}[h!]
{\includegraphics[width=0.45\textwidth,height=0.35\textwidth]{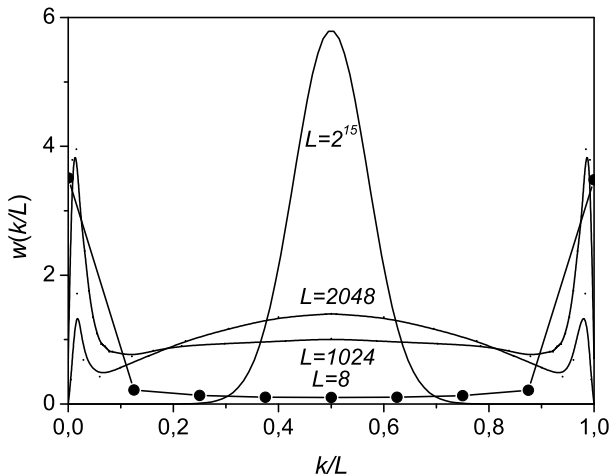}}
\caption{The distribution function $w(k/L)=LW_L(k)$ for $N$=8 and
$\delta=1/150$. Different values of the length $L$ of words is
shown near the curves.} \label{f6}
\end{figure}

The described structure of the fluctuations defines the
statical properties of the $L$-words with $L>N$ in the case
of strong persistence. The distribution function differs
significantly from the Gaussian and is characterized by a
concave form at $L\lesssim l_c \sim N/\delta $. As $L$
increases, the correlations between different parts of the
$L$-words get weaker and the $L$-word can be considered as
consisting of a number of independent sub-words. So,
according to the general mathematical
theorems~\cite{nag,ibr}, the distribution function takes on
the usual Gaussian form. Such an evolution of the
distribution function is depicted in Fig.~\ref{f6}.

\begin{figure}[h!]
{\includegraphics[width=0.45\textwidth,height=0.35\textwidth]{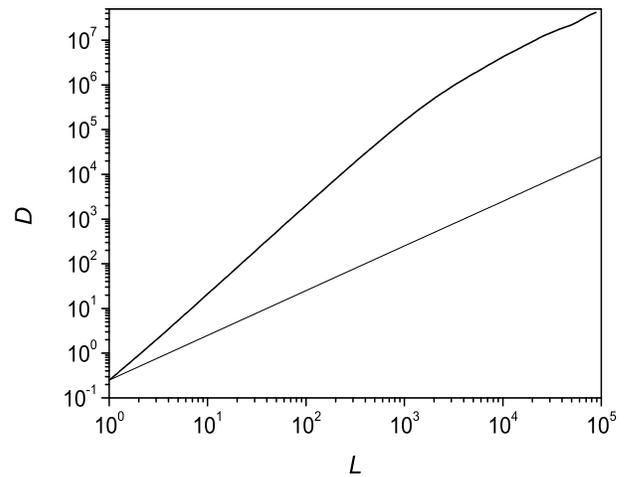}}
\caption{The dependence of the variance $D$ on the word length $L$
for the sequence with $N=20$ and $\mu=50/101$ (solid line). The
thin solid line describes the non-correlated Brownian diffusion,
$D(L)=L/4.$ } \label{f7}
\end{figure}

The variance $D(L)$ follows the quadratic law $D=L^2/4$ (see
Eq.~(\ref{45c})) up to the range of $L\lesssim l_c \sim N/\delta$
and then approaches to the asymptotics $D(L) = B L$ with $B \sim
N/4\delta$ (see Fig.~\ref{f7}).

The upper curve in Fig.~\ref{f4} presents the correlation
function for the case of strong persistence ($\mu = 50/101$,
$N=20$).

\section{Results of
numerical simulations and applications}

In this section, we support the obtained analytical results
by numerical simulations of the Markov chain with the
conditional probability Eq.~(\ref{14}). Besides, the
properties of the studied binary $N$-step Markov chain are
compared with those for the natural objects, specifically
for the coarse-grained written and DNA texts.

\subsection{Numerical simulations of the Markov chain}

The first stage of the construction of the $N$-step Markov chain
was a generation of the initial non-cor\-re\-la\-ted $N$ symbols,
zeros and unities, identically distributed with equal
probabilities 1/2. Each consequent symbol was then added to the
chain with the conditional probability determined by the previous
$N$ symbols in accordance with Eq.~(\ref{14}). Then we numerically
calculated the variance $D(L)$ by means of Eq.~(\ref{7}). The
circles in Fig.~\ref{f5} represent the calculated variance $D(L)$
for the case of weak persistence ($n=12.5 \gg 1$). A very good
agreement between the analytical result (\ref{56}) and the
numerical simulation can be observed. The case of strong
persistence is illustrated by Figs.~\ref{f6} and \ref{f7} where
the distribution function $W_L(k)$ and the variance $D(L)$ are
calculated numerically for $n=4/37$ and $n=0.1$, respectively. The
dots on the curves in Fig.~\ref{f4} represent the calculated
results for the correlation function ${\cal K}(r)$ for $n=0.1$
(the upper curve) and $n=40$ (the lower curve).

The numerical simulation was also used for the demonstration of
the proposition $\spadesuit$ (Fig.~1) and the self-similarity
property of the Markov sequence (Fig.~3). The squares in Fig.~3
represent the variance $D(L)$ for the sequence obtained by the
stochastic decimation of the initial Markov chain (solid line)
where each symbol was omitted with the probability 1/2. The
circles in this figure correspond to the regular reduction of the
sequence by removing each second symbol.

And finally, the numerical simulations have allowed us to
make sure that we are able to determine the parameters $N$
and $\mu$ of a given binary sequence. We generated the
Markov sequences with different parameters $N$ and $\mu$ and
defined numerically the corresponding curves $D(L)$. Then we
solved the inverse problem of the reconstruction of the
parameters $N$ and $\mu$ by analyzing the curves $D(L)$. The
reconstructed parameters were always in good agreement with
their prescribed values. In the following subsections we
apply this ability to the treatment of the statistical
properties of literary and DNA texts.

\subsection{Literary texts}

\begin{figure}[h!]
{\includegraphics[width=0.45\textwidth,height=0.35\textwidth]{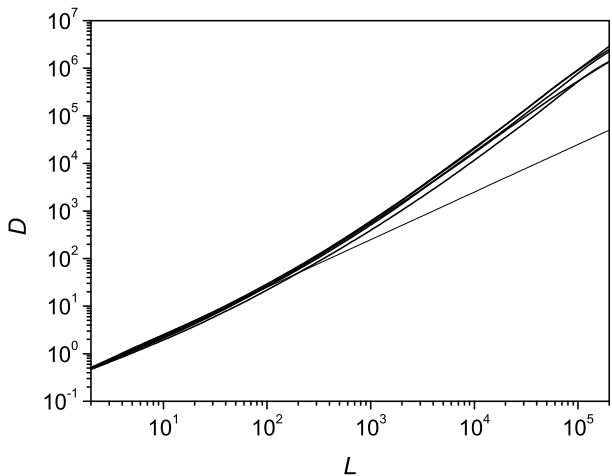}}
\caption{The dependence $D(L)$ for the coarse-grained text of
Bible obtained by means of five different kinds of random
mapping.} \label{f8}
\end{figure}

It is well-known that the statistical properties of the
coarse-grained texts written in any language exhibit a
remarkable deviation from random
sequences~\cite{schen,kant}. In order to check the
applicability of the theory of the binary $N$-step Markov
chains to literary texts we resorted to the procedure of
coarse graining by the random mapping of all characters of
the text onto the binary set of symbols, zeros and unities.
The statistical properties of the coarse-grained texts
depend, but not significantly, on the kind of mapping. This
is illustrated by the curves in Fig.~\ref{f8} where the
variance $D(L)$ for five different kinds of the mapping of
Bible is presented. In general, the random mapping leads to
nonequal numbers of unities and zeros, $k_1$ and $k_0$, in
the coarse-grained sequence. A particular analysis indicates
that the variance $D(L)$ (\ref{32}) gets the additional
multiplier,
\[
\frac{4k_0 k_1}{(k_0+k_1)^2},
\]
in this biased case. In order to derive the function $D(L)$ for
the non-biased sequence, we divided the numerically calculated
value of the variance by this multiplier.

\begin{figure}[h!]
{\includegraphics[width=0.45\textwidth,height=0.35\textwidth]{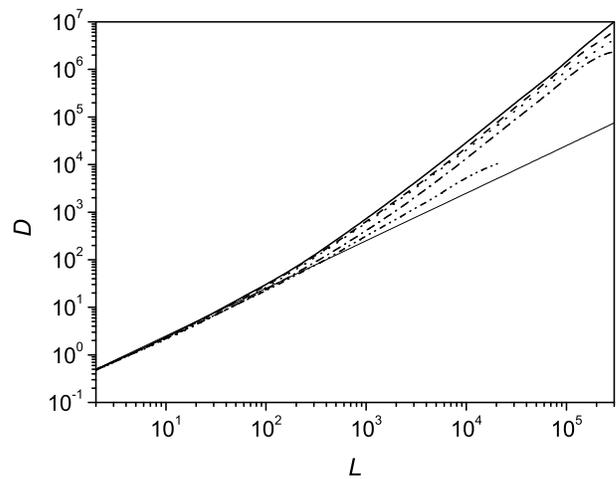}}
\caption{The dependence $D(L)$ for the coarse-grained texts of
collection of works on the computer science ($m=2.2\cdot 10^{-3}$,
solid line), Bible in Russian ($m=1.9\cdot 10^{-3}$, dashed line),
Bible in English ($m=1.5\cdot 10^{-3}$, dotted line), ``History of
Russians in the 20-th Century'' by Oleg Platonov ($m=6.4\cdot
10^{-4}$, dash-dotted line), and ``Alice's Adventures in
Wonderland'' by Lewis Carroll ($m=2.7\cdot 10^{-4}$,
dash-dot-dotted line).} \label{f9}
\end{figure}

\begin{figure}[h!]
{\includegraphics[width=0.45\textwidth,height=0.35\textwidth]{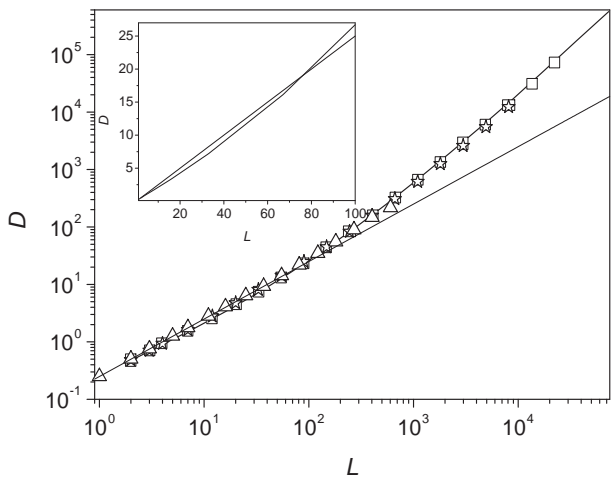}}
\caption{The dependence of the variance $D$ on the tuple length
$L$ for the coarse-grained text of Bible (solid line) and for the
decimated sequences with different parameters $\lambda$: $\lambda
= 3/4 $ (squares), $\lambda = 1/2 $ (stars), and $\lambda = 1/256
$ (triangles). The insert demonstrates the anti-persistent portion
of the $D(L)$ plot for Bible.}\label{f10}
\end{figure}

The study of different written texts has suggested that all
of them are featured by the pronounced persistent
correlations. It is demonstrated by Fig.~\ref{f9} where the
five variance curves go significantly higher than the
straight line $D=L/4$. However, it should be emphasized that
regardless of the kind of mapping the initial portions,
$L<80$, of the curves correspond to a slight anti-persistent
behavior (see insert to Fig.~\ref{f10}). Moreover, for some
inappropriate kinds of mapping (e.g., when all vowels are
mapped onto the same symbol) the anti-persistent portions
can reach the values of $L\sim 1000$. To avoid this problem,
all the curves in Fig.~\ref{f9} are obtained for the
definite representative mapping: (a-m) $\rightarrow$ 0;
(n-z) $\rightarrow$ 1.

Thus, the persistence is the common property of the binary
$N$-step Markov chains that have been considered in this
paper and the coarse-grained written texts at large scales.
Moreover, the written texts as well as the Markov sequences
possess the property of the self-similarity. Indeed, the
curves in Fig.~\ref{f10} obtained from the text of Bible
with different levels of the deterministic decimation
demonstrate the self-similarity. Presumably, this property
is the mathematical reflection of the well-known hierarchy
in the linguistics: \textit{letters $\rightarrow$ syllables
$\rightarrow$ words $\rightarrow$ sentences $\rightarrow$
paragraphs $\rightarrow$ chapters $\rightarrow$ books}.

All the above-mentioned circumstances allow us to suppose
that our theory of the binary $N$-step Markov chains can be
applied to the description of the statistical properties of
the texts of natural languages. However, in contrast to the
generated Markov sequence (see Fig.~\ref{f5}) where the full
length $\mathcal{M}$ of the chain is far greater than the
memory length $N$, the coarse-grained texts described by
Fig.~\ref{f9} are of relatively short length
$\mathcal{M}\lesssim N$. In other words, the coarse-grained
texts are similar not to the Markov chains but rather to
some non-stationary short fragments. This implies that each
of the written texts is correlated throughout the whole of
its length. Therefore, as fae as the written texts are
concerned, it is impossible to observe the second portion of
the curve $D(L)$ parallel (in the log-log scale) to the line
$D(L)=L/4$, similar to that shown in Fig.~\ref{f5}. As a
result, one cannot define the values of both parameters $N$
and $\mu$ for the coarse-grained texts. The analysis of the
curves in Fig.~6 can give the combination $m=2\mu/N(1-2\mu)$
only (see Eq.~(\ref{32})). Perhaps, this particular
combination is the real parameter governing the persistent
properties of the literary texts.

We would like to note that the origin of the long-range
correlations in the literary texts is hardly related to the
grammatical rules as is claimed in Ref.~\cite{kant}. At
short scales $L\leq 80$ where the grammatical rules are in
fact applicable the character of correlations is
anti-persistent (see the insert in Fig.~\ref{10}) whereas
semantic correlations lead to the global persistent behavior
of the variance $D(L)$ throughout the entire of literary
text.

The numerical estimations of the persistent parameter $m$
and the characterization of the languages and different
authors using this parameter can be regarded as a new
intriguing problem of linguistics. For instance, the
unprecedented low value of $m$ for the very inventive work
by Lewis Carroll as well as the closeness of $m$ for the
texts of English and Russian versions of Bible are of
certain interest.

It should be noted that there exist special kinds of short-range
correlated texts which can be specified by both of the parameters,
$N$ and $\mu$. For example, all dictionaries consist of the
families of words where some preferable letters are repeated more
frequently than in their other parts. Yet another example of the
shortly correlated texts is any lexicographically ordered list of
words. The analysis of written texts of this kind is given below.

\subsection{Dictionaries}

\begin{figure}[h!]
{\includegraphics[width=0.45\textwidth,height=0.35\textwidth]{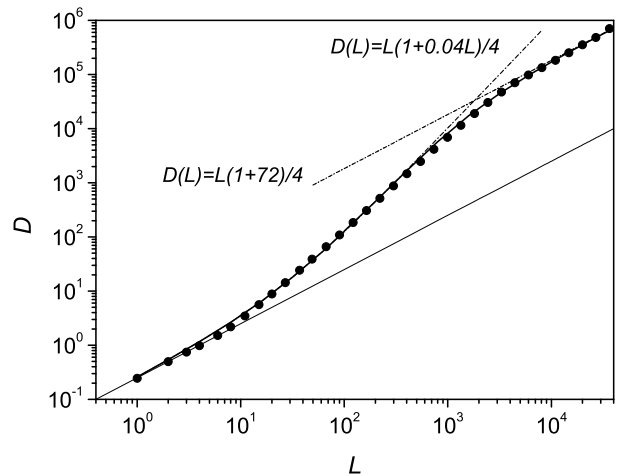}}
\caption{The dependence $D(L)$ for the coarse-grained alphabetical
list of 15462 English words (circles). The solid line is the plot
of function Eq.~(55) with the fitting parameters $N=180$ and
$\mu=0.44$.} \label{f11}
\end{figure}

As an example, we have investigated the statistical
properties of the coarse-grained alphabetical
(lexicographically ordered) list of the most frequently used
15462 English words. In contrast to other texts, the
statistical properties of the coarse-grained dictionaries
are very sensitive to the kind of mapping. If one uses the
above-mentioned mapping, (a-m) $\rightarrow$ 0; (n-z)
$\rightarrow$ 1, the behavior of the variance $D(L)$ similar
to that shown in Fig.~\ref{f9} would be obtained. The
particular construction of the dictionary manifests itself
if the preferable letters in the neighboring families of
words are mapped onto the different symbols. The variance
$D(L)$ for the dictionary coarse-grained by means of such
mapping is shown by circles in Fig.~\ref{f11}. It is clearly
seen that the graph of the function $D(L)$ consists of two
portions similarly to the curve in Fig.~\ref{f5} obtained
for the generated $N$-step Markov sequence. The fitting of
the curve in Fig.~\ref{f11} by function (\ref{56}) (solid
line in Fig.~\ref{f11}) yielded the values of the parameters
$N=180$ and $\mu =0.44$.

\subsection{DNA texts}

It is known that any DNA text is written by four ``characters'',
specifically by adenine (A), cytosine (C), guanine (G), and
thymine (T). Therefore, there are three nonequivalent types of the
DNA text mapping onto one-dimensional binary sequences of zeros
and unities. The first of them is the so-called purine-pyrimidine
rule, \{A,G\} $\rightarrow$ 0, \{C,T\} $\rightarrow$ 1. The second
one is the hydrogen-bond rule, \{A,T\} $\rightarrow$ 0, \{C,G\}
$\rightarrow$ 1. And, finally, the third is \{A,C\} $\rightarrow$
0, \{G,T\} $\rightarrow$ 1.

\begin{figure}[h!]
{\includegraphics[width=0.45\textwidth,height=0.35\textwidth]{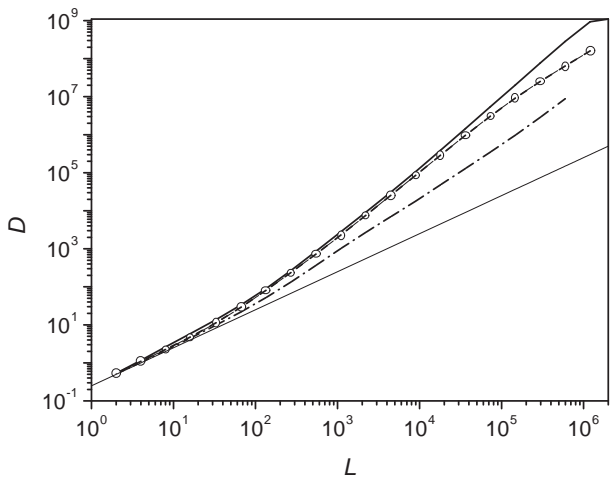}}
\caption{The dependence $D(L)$ for the coarse-grained DNA
text of \textit{Bacillus subtilis, complete genome}, for
three nonequivalent kinds of mapping. Solid, dashed, and
dash-dotted lines correspond to the mappings \{A,G\}
$\rightarrow$ 0, \{C,T\} $\rightarrow$ 1 (the parameter
$m=4.1\cdot 10^{-2}$), \{A,T\} $\rightarrow$ 0, \{C,G\}
$\rightarrow$ 1 ($m=2.5\cdot 10^{-2}$), and \{A,C\}
$\rightarrow$ 0, \{G,T\} $\rightarrow$ 1 ($m=1.5\cdot
10^{-2}$), respectively.} \label{f12}
\end{figure}

By\hfill way\hfill of\hfill example,\hfill the\hfill
variance\hfill $D(L)$\hfill for\hfill the\hfill coar-
\\se-grained\hfill
text\hfill of\hfill \textit{Bacillus\hfill subtilis,\hfill
complete\hfill genome} \\
(ftp:$//$ftp.ncbi.nih.gov$/$genomes$/$bacteria$/$bacillus\_subti-
\\lis$/$NC\_000964.gbk) is displayed in Fig.~\ref{f12} for all
possible types of mapping. One can see that the persistent
properties of DNA are more pronounced than for the written texts
and, contrary to the written texts, the $D(L)$ dependence for DNA
does not exhibit the anti-persistent behavior at small values of
$L$. In our view, the noticeable deviation of different curves in
Fig.~\ref{f12} from each other demonstrates that the DNA texts are
much more complex objects in comparison with the written ones.
Indeed, the different kinds of mapping reveal and emphasize
various types of physical attractive correlations between the
nucleotides in DNA, such as the strong purine-purine and
pyrimidine-pyrimidine persistent correlations (the upper curve),
and the correlations caused by a weaker attraction
A$\leftrightarrow$T and C$\leftrightarrow$G (the middle curve).

\begin{figure}[h!]
{\includegraphics[width=0.45\textwidth,height=0.35\textwidth]{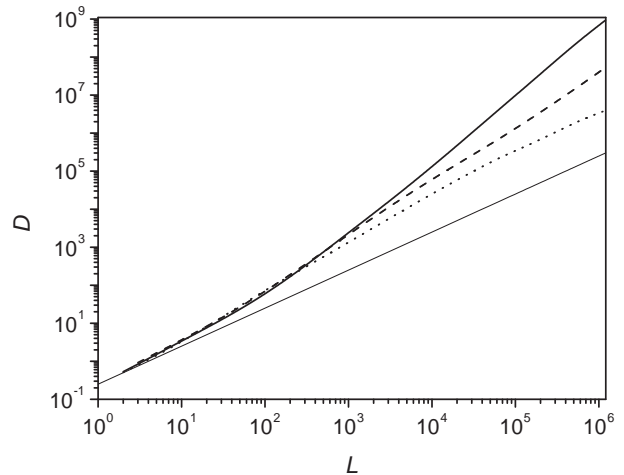}}
\caption{The dependence $D(L)$ for the coarse-grained DNA texts of
\textit{Bacillus subtilis, complete genome}, the Bacteria, (solid
line); \textit{Methanosarcina acetivorans, complete genome}, the
Archaea, (dashed line); \textit{Drosophila melanogaster chromosome
X, complete sequence}, the Eukarya, (dotted line) for the mapping
\{A,G\} $\rightarrow$ 0, \{C,T\} $\rightarrow$ 1.} \label{f13}
\end{figure}

It is interesting to compare the correlation properties of
the DNA texts for three different species that belong to the
major domains of living organisms: the Bacteria, the
Archaea, and the Eukarya~\cite{mad}. Figure~\ref{f13} shows
the variance $D(L)$ for the coarse-grained DNA texts of
Bacillus subtilis (the Bacteria), Methanosarcina
acetivoransthe (the Archaea), and Drosophila melanogaster -
fruit fly - (the Eukarya) for the most representative
mapping \{A,G\} $\rightarrow$ 0, \{C,T\} $\rightarrow$ 1. It
is seen that the $D(L)$ curve for the DNA text of Bacillus
subtilis is characterized by the highest persistence. As
well as for the written texts, the $D(L)$ curves for the DNA
of both the Bacteria and the Archaea do not contain the
linear portions given by Eq.~(\ref{58}). This suggests that
their DNA chains are not stationary sequences. In this
connection, we would like to point out that their DNA
molecules are circular and represent the collection of
extended coding regions interrupted by small non-coding
regions. According to Figs.~\ref{f12}, \ref{f13}, the
non-coding regions do not disrupts the correlation between
the coding areas, and the DNA systems of the Bacteria and
the Archaea are fully correlated throughout their entire
lengths. Contrary to them, the DNA molecules of the Eukarya
have the linear structure and contain long non-coding
portions. As evident from Fig.~\ref{f13}, the DNA sequence
of the representative of the Eukarya is not entirely
correlated. The $D(L)$ curve for the X-chromosome of the
fruit fly corresponds qualitatively to Eqs.~(\ref{56}),
(\ref{57}) with $\mu \approx 0.35$ and $N \approx 250$. If
one draws an analogy between the DNA sequences and the
literary texts, the resemblance of the correlation
properties of integral literary novels and the DNA texts of
the Bacteria and Archaea are to be found, whereas the DNA
texts of the Eukarya are more similar to the collections of
10$^4$--10$^5$ short stories.

\section{Conclusion}

Thus, we have developed a new approach to describing the
strongly correlated one-dimensional systems. The simple,
exactly solvable model of the uniform binary $N$-step Markov
chain is presented. The memory length $N$ and the parameter
$\mu$ of the persistent correlations are two parameters in
our theory. The correlation function ${\cal K}(r)$ is
usually employed as the input characteristics for the
description of the correlated random systems. Yet, the
function ${\cal K}(r)$ describes not only the direct
interconnection of the elements $a_i$ and $a_{i+r}$, but
also takes into account their indirect interaction via other
elements. Since our approach operates with the ``original''
parameters $N$ and $\mu$, we believe that it allows us to
reveal the intrinsic properties of the system which provide
the correlations between the elements.

We have demonstrated the applicability of the developed
theoretical model to the different kinds of relatively
weakly correlated stochastic systems. Perhaps, the case of
strong persistency is also of interest from the standpoint
of possible applications. Indeed, the domain structure of
the symbol fluctuations at $n\ll 1$ is very similar to the
domains in magnetics. Thus, an attempt to model the magnetic
structures by the Markov chains with strongly pronounced
persistent properties can be appropriate.

We would like to note that there exist some features of the
real correlated systems which cannot be interpreted in terms
of our two-parametric model. For example, the interference
of the grammatical anti-persistent and semantic persistent
correlations in the literary texts requires more than two
parameters for their description. Obviously, more complex
models should be worked out for the adequate interpretation
of the statistical properties of the DNA texts and other
real correlated systems. In particular, the Markov chains
consisting of more than two different elements (non-binary
chains) can be suitable for modelling the DNA systems.

\acknowledgments We acknowledge to Yu.~L.~Rybalko and
A.~L.~Patsenker for the assistance in the numerical simulations,
M.~E.~ Serbin and R.~Zomorrody for the helpful discussions.

\end{document}